\documentclass[prb,twocolumn,showpacs,superscriptaddress]{revtex4}%
\usepackage{amsfonts}
\usepackage{amsmath}
\usepackage{amssymb}
\usepackage{longtable}
\usepackage{graphicx}%
\providecommand{\U}[1]{\protect\rule{.1in}{.1in}}

\begin{document}
\title{Quantum molecular dynamic simulations of warm dense carbon monoxide}
\author{Yujuan Zhang}
\affiliation{LCP, Institute of Applied Physics and Computational
Mathematics, P.O. Box 8009, Beijing 100088, People's Republic of
China}
\author{Cong Wang}
\affiliation{LCP, Institute of Applied Physics and Computational
Mathematics, P.O. Box 8009, Beijing 100088, People's Republic of
China}
\author{Dafang Li}
\affiliation{LCP, Institute of Applied Physics and Computational
Mathematics, P.O. Box 8009, Beijing 100088, People's Republic of
China}
\author{Ping Zhang}
\thanks{Corresponding author: zhang\underline{ }ping@iapcm.ac.cn}
\affiliation{LCP, Institute of Applied Physics and Computational
Mathematics, P.O. Box 8009, Beijing 100088, People's Republic of
China} \affiliation{Center for Applied Physics and Technology,
Peking University, Beijing 100871, People's Republic of China}

\begin{abstract}
Using quantum molecular dynamic simulations, we have studied the
thermophysical properties of warm dense carbon monoxide under
extreme conditions. The principal Hugoniot, which is derived from
the equation of state, shows excellent agreement with available
experimental data up to 67 GPa. The chemical decomposition of
carbon monoxide has been predicted at 8 GPa by means of pair
correlation function. Based on Kubo-Greenwood formula, the dc
electrical conductivity and the optical reflectivity are
determined, and the nonmetal-metal transition for shock compressed
carbon monoxide is observed around 43 GPa.
\end{abstract}
\maketitle

\section{INTRODUCTION}
With rapid development of planetary science and explosion
technology, pressure induced responses of materials under extreme
conditions, where the combination of high temperatures and high
pressures defines the warm dense regime, have recently attracted
extensive attention \cite{Erns2009,Heml2000,Maze2004}.
Experimental techniques, e.g. gas gun \cite{Nell1981}, chemical
explosives \cite{Mint2006}, pulsed power \cite{Knud2003,Boch2011},
have been adopted in dynamic compressions, where pressures could
reach megabar region. Due to the enormous progress in
computational capacity, quantum molecular dynamics (QMD)
\cite{Kress2004,Hols2008}, provide powerful tools to study warm
dense matter theoretically, where dissociations, recombinations,
and ionizations characterize the high pressure behaviors.

As one of the major components in the reacted high explosives, the
pressure introduced transitions of carbon monoxide (CO) are of
crucial interest in exploring detonation products
\cite{Kerl1980,Ross1980,Schm1997}. The equation of states (EOS) and
the relative quantities, for instance the Hugoniot points, sound
velocity and particle velocity, are important parameters for
materials under extreme conditions. Furthermore, electrical and
optical properties such as, e.g., dc conductivity ($\sigma_{dc}$)
and optical reflectance are closely related to the dielectric
function, which can be evaluated through dynamic conductivity. The
above mentioned quantities are of crucial interest in characterizing
the unique behavior of warm dense CO. The Hugoniot points up to 70
GPa have been detected by a two-stage light-gas gun \cite{Nell1981},
and three theoretical models were used to analyze the EOS and
chemical decomposition. CO was reported to decomposes into condensed
carbon and fluid CO$_2$ at high pressures by pair potential
calculations \cite{Ross1980}, and the dissociation rates were then
determined \cite{Alia2003}. The principal Hugoniot and decomposition
of CO have been investigated by means of QMD method
\cite{Kress2004}. However, wide range EOS, shock products,
electrical and optical properties of CO under extreme conditions are
still yet to be presented and understood.

In the present work, we have performed QMD simulations to study
the thermophysical properties of warm dense CO. The decomposition
and recombination of CO along the principal Hugoniot are analyzed
by pair correlation functions (PCF). The dynamic conductivity is
calculated by Kubo-Greenwood formula, from which the dc
conductivity, the dielectric function and optical reflectivity can
be extracted. The rest of the paper is arranged as follows. In
section II, the computational method is briefly described. We
present and discuss our calculated results, some of which are made
comparison with available experimental and theoretical results in
section III. Finally, we close our paper with a summary of our
main results.

\section{COMPUTATIONAL METHOD}

In this study, the molecular dynamics trajectories are calculated by
employing the Vienna $ab$ $initio$ simulation package (VASP)
plane-wave pseudopotential code developed at the Technical
University of Vienna \cite{Kres1993,Kres1996}. The calculations are
performed by a series of volume-fixed supercell with an invariable
number of atoms. Employing the Born-Oppenheimer approximation, the
electrons are fully quantum-mechanical treated by solving the
Kohn-Sham equations for the orbitals and energies within a
plane-wave, finite-temperature DFT formulation \cite{Leno2000},
where the electronic states are populated according to the
Fermi-Dirac distribution at electron temperature $T$$_e$. The
Perdew-Wang 91 (PW91) generalized gradient approximation
\cite{Perd1991} and the projector augmented wave (PAW) potential of
Bl\"{o}chl \cite{Bloc1994} are employed to describe the
exchange-correlation energy and the electron-ion interaction,
respectively. The isokinetic ensemble (NVT) is employed for the
ions, where the ionic temperature $T$$_i$ is controlled by No\'{s}e
thermostat \cite{Nose1984} and the system is controlled in local
thermodynamical equilibrium by setting the electron temperature
$T$$_e$ and the ion temperature $T$$_i$ to be equal.

The supercell in our calculation contains 128 atoms (64 CO
molecules) with periodic boundary condition. The plane-wave cutoff
energy is selected to be 550 eV, which is tested to be good
convergence for both total energy and pressure. In molecular
dynamic simulation, only $\Gamma$ point of the Brillouin zone is
included, while, 4$\times$4$\times$4 Monkhorst-Pack scheme
\cite{Monk1976} grid points are used in the electronic structure
calculations. The selected densities range from 0.807 to 3.10
g/cm$^3$ with temperatures from 77 to 11000 K, which highlight the
principal shock Hugoniot region. All the dynamic simulations last
3-5 ps with time steps of 0.5-1.0 fs according to different
conditions. The EOS data and pair correlation functions are
obtained by averaging over all the particles and the final 1-2 ps
simulations.

\section{results and discussion}
\subsection{Equation of state and pair correlation function}
\begin{table}[ptb]
\caption{Hugoniot pressure ($P$) and temperature ($T$) points
derived from QMD simulations at a series of densities ($\rho$).
The corresponding particle velocity ($u$$_p$),
and shock velocity ($u$$_s$) are also displayed.}%
\begin{ruledtabular}
\renewcommand{\tabcolsep}{0.01pc}
\begin{tabular}{ccccccc}
$\rho$&$P$&$T$&$u$$_p$&$u$$_s$\\
(g/cm$^3$)&(GPa)&(K)&(km/s)&(km/s)\\
\hline
1.45&5.32&603&1.70&3.83\\
1.78&8.57&1614&2.40&4.39\\
2.25&20.20&4545&4.00&6.23\\
2.52&31.77&6600&5.17&7.60\\
2.70&43.27&8697&6.12&8.74\\
2.90&53.30&9175&6.90&9.56\\
3.10&67.27&11000&7.84&10.61\\
\end{tabular}
\label{bader}
\end{ruledtabular}
\end{table}

A precise description of EOS plays an important role in accurately
calculating the electrical and optical properties. The EOS is
examined theoretically along the Hugoniot, which can be derived
from conservation of mass, momentum, and energy for a isolated
system compressed by a pusher at a constant velocity.
Rankine-Hugoniot equation describes the shock adiabat through a
relation between the initial and final internal energies,
pressures and volumes as follows \cite{Zeld1966}:
\begin{eqnarray}
\left(E_{1}-E_{0}\right)+\frac{1}{2}\left(V_{1}-V_{0}\right)\left(P_{0}+P_{1}\right)
& = & 0,
\end{eqnarray}
where $E$, $V$ and $P$ are the internal energy, volume, and
pressure, and the subscripts 0 and 1 denote the initial and
shocked states, respectively. In our canonical ensemble
calculations, the internal energy $E$ equals to the sum of the
total energy from the finite-temperature DFT calculation and
zero-point energy. The pressure $P$ consists of the electronic
$P_{e}$ and ionic $P_{i}$ components. The pressure can be obtained
by $P=P_{e}+\rho_{n}k_{B}T$, where $\rho_{n}$ is the ion number
density, and $k$$_B$ is the Boltzman constant.

\begin{figure}
\includegraphics[width=0.85\linewidth]{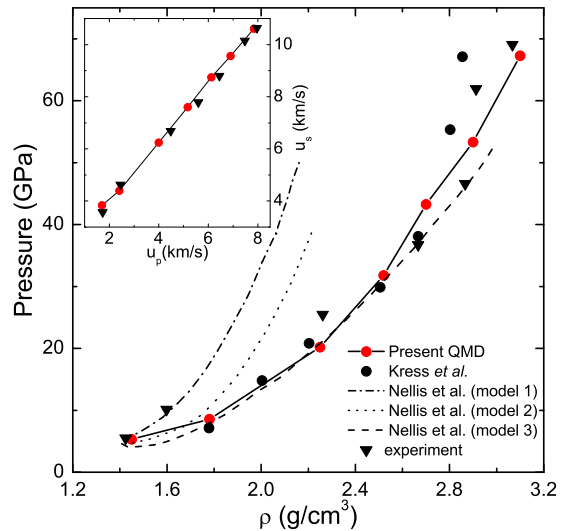}
\caption{(Color online) Principal Hugoniot (the main panel) and
the ($u$$_s$, $u$$_p$) diagram (the inset) of CO. For comparison,
the present QMD results, together with the previous experimental
data (Ref. 4) and other theoretical results (Ref. 4 and Ref. 12)
are all provided.}
\end{figure}
As the starting point along the Hugoniot, the initial density is
$\rho$$_{0}$=0.807 g/cm$^{3}$ at a temperature of 77 K
\cite{Kress2004}, with the relative internal energy $E_{0}=-14.84$
eV/molecule. The initial pressure can be neglected compared to the
high pressure of the final state. The Hugoniot points are
determined in the following way. For a given density $\rho$, a
series of simulations are executed for different temperature $T$.
$(E_{1}-E_{0})$ and $(V_{0}-V_{1})(P_{0}+P_{1})/2$ as a function
of temperature are then fitted to polynomial expansions of $T$.
The principal Hugoniot temperature $T_{1}$ and pressure $P_{1}$
can be determined by solving Eq. (1). The particle velocity
$u_{p}$ and shock velocity $u_{s}$ are then determined from the
other two Rankine-Hugoniot equations \cite{Zeld1966},
$V_{1}=V_{0}\left[1-\left(u_{p}/u_{s}\right)\right]$ and
$P_{1}-P_{0}=\rho_{0}u_{s}u_{p}$. The principal Hugoniot points of
CO derived from Eq. (1) as well as the particle velocity $u_{p}$
and shock velocity $u_{s}$ are listed in Table I.

The simulated principal Hugoniot points for CO are shown in Fig.
1, where previous experimental and theoretical data are also
provided for comparison. As seen in Fig. 1, our simulated results
accord excellently with the experimental results \cite{Nell1981}
and the theoretical results \cite{Kress2004}. The ($u_{p}$,
$u_{s}$) diagram also shows good agreement with experimental
results (the inset of Fig. 1).

\begin{figure}
\includegraphics[width=0.90\linewidth]{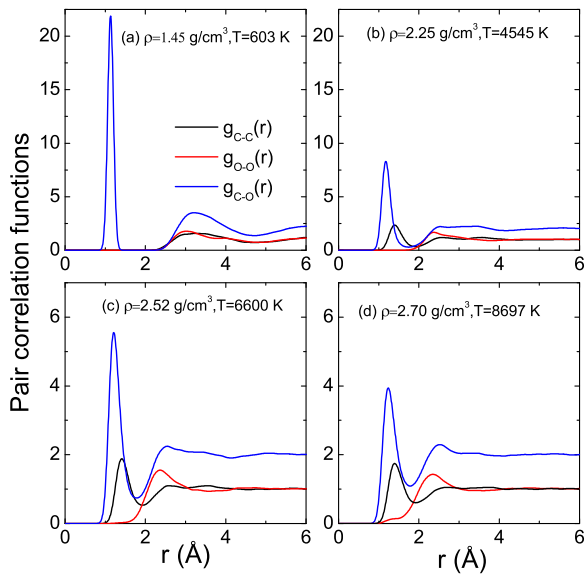}
\caption{(Color online) Pair-correlation functions for C--C (black
line), O--O (red line), C--O (blue line) along the principal
Hugoniot of CO.}
\end{figure}

To examine the structural transitions of CO under shock
compressions, the PCFs
[$g(r)$=$\frac{1}{<\rho>}$$\frac{dn(R,R+dR)}{dv(R,R+dR)}$], which
represent the possibility of finding a particle at a distance $r$
from a reference atom, are calculated. The PCFs and the
corresponding atomic structures along the principal Hugoniot of
CO, are presented in Fig. 2 and Fig. 3, respectively. At the
pressure of 5 GPa ($\rho$=1.45 g/cm$^{3}$ and $T$=603 K), the peak
of C--O PCF g$_{\rm{C-O}}$($r$) locates around 1.13
$\textrm{\AA}$, which corresponds to the equilibrium internuclear
distance of the C--O bond in CO molecule \cite{Gill1950} [Fig.
2(a)], indicating that CO molecules remain their ideal molecular
fluid state without dissociation. From the corresponding atomic
structure [Fig. 3(a)], it can be clearly seen that the supercell
is filled with CO molecules. With the increase of pressure along
the Hugoniot, the main peak of PCF g$_{\rm{C-O}}$($r$) becomes
reduced in amplitude and gets broadened due to the thermal
excitation, as shown in Fig. 2(b). Meanwhile, the
g$_{\rm{C-C}}$($r$) appears a peak at $r$=1.47\AA, which lies
between typical $sp$$^2$ and $sp$$^3$ hybrid C--C bond length.
From the atomic structure of CO at $\rho$=2.25 g/cm$^3$ ($P$=20.2
Gpa, $T$=4545 K), CO$_2$ molecules and the small clusters with
carbon backbones can be clearly observed [Fig. 3(b)]. The
structural transitions of CO may bring with the increase of the
$\sigma_{dc}$, which will be discussed in the following section.
As the difference between the length of C-O bond in CO molecule
(1.13\AA) and that of CO$_2$ molecule (1.16\AA) \cite{wang2010} is
very small, the appearances of their PCFs are similar. Despite the
fact that the formation of CO$_2$ increases the peak value of
g$_{\rm{C-O}}$($r$), CO is further dissociated leading to a
decrease in the peak value. Under the competition of the above two
effects, the amplitude of PCF g$_{\rm{C-O}}$($r$) still keeps
reduced as the pressure increases. At higher pressures, CO$_2$
molecules get dissociated, and a mixture of CO, atomic carbon, and
some carbon backbones clusters forms, which results in a further
decrease of the peak value of PCF g$_{\rm{C-O}}$($r$). In
particular, diatomic oxygen [the peak of PCF g$_{\rm{O-O}}$($r$)
is around 1.23 \AA\ as shown in Fig. 2(d)] and monoatomic oxygen
are formed in the shocked system [Fig. 3(d)], which lead to the
nonmetal-metal transition. Overall, the systematic behavior of the
dissociation and recombination of fluid CO under shock pressure is
described. At low pressures ($P$$\le$20 GPa), CO$_2$ as well as
some carbon backbone clusters forms. With increasing the pressure
($P$$\ge$43GPa), CO$_2$ dissociates and a mixture of CO, atomic
carbon diatomic oxygen monoatomic oxygen and carbon backbone
clusters forms.

\begin{figure}
\includegraphics[width=0.85\linewidth]{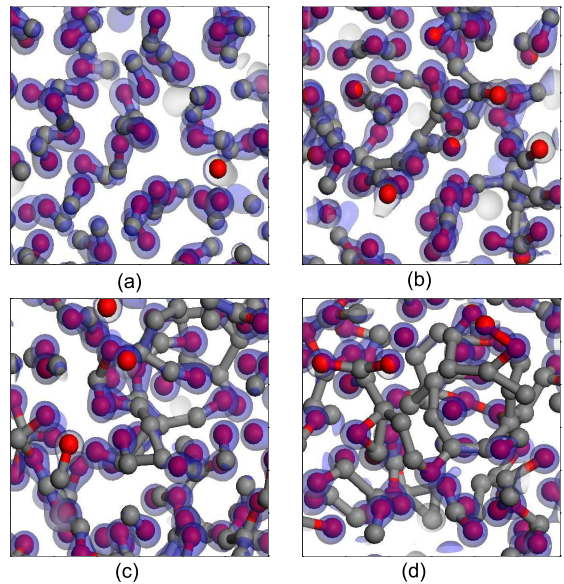}
\caption{(Color online) Atomic structures at different Hugoniot
points. The carbon and oxygen atoms are denoted by gray and red
balls, respectively, and the relative isosurface of charge density
(blue regimes) are plotted. (a)$\rho$=1.45 g/cm$^3$, $T$=603 K;
(b)$\rho$=2.25 g/cm$^3$, $T$=4545 K; (c)$\rho$=2.52 g/cm$^3$,
$T$=6600 K; (d)$\rho$=2.70 g/cm$^3$, $T$=8697 K.}
\end{figure}

\subsection{Dynamic conductivity under shock compressions}

The dissociation of molecules under dynamic compression are
closely related to the nonmetal-metal transition properties. The
dynamic conductivity is derived from the Kubo-Greenwood formula as
follows \cite{Kubo1957,Gree1958}:
\begin{eqnarray}
\sigma_{1}(\omega) & = & \frac{2\pi}{3\omega\Omega}\underset{\textbf{k}}{\sum}w(\textbf{k})\overset{N}{\underset{j=1}{\sum}}\overset{N}{\underset{i=1}{\sum}}\overset{3}{\underset{\alpha=1}{\sum}}[f(\epsilon_{i},\textbf{k})-f(\epsilon_{j},\textbf{k})]\nonumber \\
 &  & \times|\langle \Psi_{j,\textbf{k}}|\nabla_{\alpha}|\Psi_{i,\textbf{k}}\rangle |^{2}\delta(\epsilon_{j,\textbf{k}}-\epsilon_{i,\textbf{k}}-\hbar\omega),
\end{eqnarray}
where $\Psi_{i,\textbf{k}}$ is Kohn-Sham eigenstate, with
corresponding eigenvalue $\epsilon_{i,\textbf{k}}$, and occupation
number $f(\epsilon_{i},\mathbf{k})$. $\Omega$ is the volume of the
supercell. $\omega$ (\textbf{k}) is the K-point weighting factor.
The $i$ and $j$ summations range over $N$ discrete bands included
in the calculation, and $\alpha$ over the three spatial
directions.

In order to calculate the conductivity, thirty atomic configurations
are selected at equilibrium along the principal Hugoniot. The
behavior of the conductivity $\sigma_{1}$ versus energy at different
Hugoniot points are calculated (not shown). With the increase of
density and temperature along the principal Hugoniot, the main peak
increases in amplitude and gets broadened. Such change of the peak
indicates the increase of another important physical quantity, dc
conductivity, which follows the static limit
$\sigma_{\rm{dc}}=\underset{\omega\rightarrow0}{lim}\sigma_{1}\left(\omega\right)$.
The dc conductivity is extracted as shown in Fig. 4. For the
pressures below 8 GPa, $\sigma_{\rm{dc}}$ can be treated as small as
negligible, which shows the insulating nature of CO at low
pressures. At higher pressures, the dc conductivity
$\sigma_{\rm{dc}}$ of CO jumps three orders of magnitude from 8 to
67 GPa (603 to 11000 K). The nonmetal-metal transition occurs around
43 Gpa, which is accompanied by the gradual dissociation of the
molecular fluid. Such nonmetal-metal transition under shock pressure
has also been reported in fluid CO$_2$ \cite{wang2010} and O$_2$
\cite{Wangc2010}.

\begin{figure}
\includegraphics[width=0.85\linewidth]{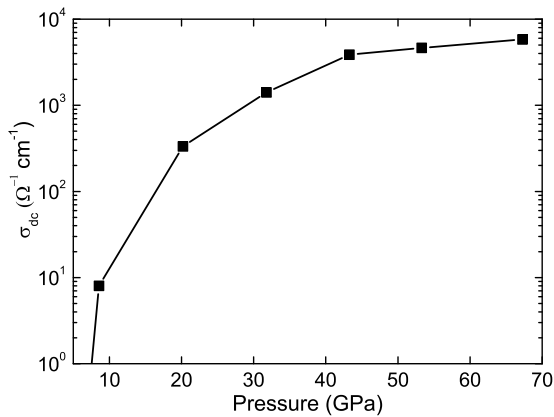}
\caption{The dc conductivity dependence on pressure is extracted
from the dynamic electrical conductivity.}
\end{figure}

The imaginary part of conductivity $\sigma_{2}$ can be obtained
via the Kramers-Kronig relation as
\begin{eqnarray}
\sigma_{2}(\omega) & = &
-\frac{2}{\pi}P\int\frac{\sigma_{1}(\nu)\omega}{(\nu^{2}-\omega^{2})}d\nu,
\end{eqnarray}
where $P$ is the principal value of the integral and $\nu$ is
frequency. The real part $\epsilon$$_{1}$($\omega$) and the
imaginary part $\epsilon$$_{2}$($\omega$) of dielectric functions
can be derived from the two parts of the conductivity,
\begin{eqnarray}
\epsilon_{1}(\omega) & = &
1-\frac{1}{\epsilon_{0}\omega}\sigma_{2}(\omega),
\end{eqnarray}
and
\begin{eqnarray}
\epsilon_{2}(\omega) & = &
\frac{1}{\epsilon_{0}\omega}\sigma_{1}(\omega).
\end{eqnarray}

The square of the index of refraction, which contains the real
part $n$($\omega$) and the imaginary part $k$($\omega$), is equal
to the dielectric function. Then the index of refraction can be
obtained by
\begin{eqnarray}
n(\omega)=\sqrt{\frac{1}{2}[|\epsilon(\omega)|+\epsilon_{1}(\omega)]},
\end{eqnarray}
and
\begin{eqnarray}
k(\omega)=\sqrt{\frac{1}{2}[|\epsilon(\omega)|-\epsilon_{1}(\omega)]}.
\end{eqnarray}
\begin{figure}
\includegraphics[width=0.85\linewidth]{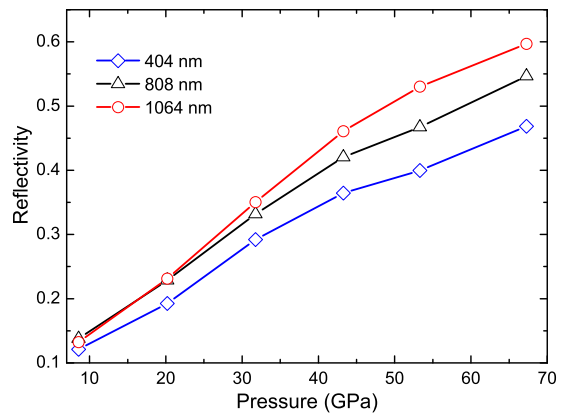}
\caption{(Color online) Optical reflectivity of shocked CO for
wavelengths of 404, 808 and 1064 nm along the principal Hugoniot.}
\end{figure}
The index of refraction is useful for evaluating optical
properties, such as the reflectivity and absorption coefficient.
Emissivity of spectrum is an important quantity, which can be used
to determine temperature in experiments. The optical reflectivity
and emissivity of spectrum are closely related. We calculated the
reflectivity of CO under shock pressure by the following relation
\begin{eqnarray}
r(\omega) & = &
\frac{[1-n(\omega)]^{2}+k(\omega)^{2}}{[1+n(\omega)]^{2}+k(\omega)^{2}}.
\end{eqnarray}
The optical reflectivities of shocked CO for wavelengths of 404, 808
and 1064 nm along the principal Hugoniot are shown in Fig. 5. The
change induced by the pressure can be clearly seen. The optical
reflectance increases from 0.12 to 0.46-0.60 with the pressure
increasing from 8 to 67 GPa, which can be interpreted as a gradual
transition from nonmetal state to metal state. Similar behavior in
the optical reflectance of liquid deuterium has been previously
observed in experiment \cite{Cell2000}. Our predictions of CO need
be tested in the future experiments.

\section{CONCLUSIONS}

In summary, we have performed QMD simulations to study the
thermophysical and optical properties of shock-compressed carbon
monoxide along the principal Hugoniot. The principal Hugoniot
curve has been derived from the EOS, which shows excellent
agreement with experimental and theoretical results. The
dissociation of CO under shock compressions is described by the
pair correlation functions. At low pressures ($P$$\le$20 GPa),
CO$_2$ and large carbon backbone clusters form. Above 43 GPa,
CO$_2$ molecules dissociate and a mixture of CO, atomic carbon and
carbon backbones clusters form. Based on the calculation of the
dynamic conductivity, the nonmetal-metal transition and the change
in optical reflectivity are observed. In particular, the
nonmetal-metal transition occurs around 43 GPa.

\begin{acknowledgments}
This work was supported by NSFC under Grants No. 11005012 and No.
51071032, by the National Basic Security Research Program of
China, and by the National High-Tech ICF Committee of China.
\end{acknowledgments}

\end{document}